# Nanosecond laser treatment of graphene

Valter Kiisk[*], Tauno Kahro, Jekaterina Kozlova, Leonard Matisen, Harry Alles

*Institute of Physics, University of Tartu, Riia Str. 142, 51014 Tartu, Estonia*

**Abstract**

Laser processing of graphene is of great interest for cutting, patterning and structural engineering purposes. Tunable nanosecond lasers have the advantage of being relatively widespread (compared to e.g. femtosecond or high-power continuous wave lasers). Hereby we have conducted an investigation of the impact of nanosecond laser pulses on CVD graphene. The damage produced by sufficiently strong single shots (pulse width 5 ns, wavelength 532 or 266 nm) from tunable optical parametric oscillator was investigated by the methods of scanning electron microscopy and optical microspectroscopy (Raman and fluorescence). Threshold of energy density for producing visible damage was found to be ~200 mJ/cm$^2$. For UV irradiation the threshold could be notably less depending on the origin of sample. Surprisingly strong fluorescence signal was recorded from damaged areas and is attributed to the residues of oxidized graphene.

*KEYWORDS*: CVD graphene; laser irradiation; photoluminescence; Raman spectroscopy

*PACS*: 81.05.ue, 79.20.Eb, 82.80.Gk

## 1. Introduction

The experimental realization of monolayer graphite sheet called graphene has led to an intense investigation of the fundamental properties and potential applications of this new carbon nanomaterial during the last decade [1]. In particular, graphene is expected to have promising optical and optoelectronic applications considering its good transparency which is accompanied with excellent electrical conductivity, saturable absorption and possibility to induce intense luminescence after special treatment [2]. For fast and high-power optical processes the high optical damage threshold is an important characteristic. On the other hand, pulsed lasers can be beneficial for special processing of graphene, like cutting and patterning with sub-micron resolution. Properly refined laser treatment can also induce more subtle chemical or structural changes like bandgap tailoring and formation of conductive channels through localized reduction of graphene oxide [3,4,5] as well as superhydrophobicity and iridescence through biomimetic structuration [6].

[*] Corresponding author. E-mail: kiisk@ut.ee; Tel.: +372-7374613; Fax: +372-7383033.

Impact of continuous wave (CW) [7] and especially femtosecond [8,9,10] laser irradiation on graphene has been characterized to some extent. While CW and femtosecond laser beams represent two extreme (and convenient) cases in terms of the involved radiation-matter interaction, lasers operating in nanosecond regime are somewhat more common and provide a wider choice of wavelengths. Recently, relatively high-power nanosecond lasers have been employed in two-beam interference geometry for large-area patterning of grapheme oxide for implementation of a gas sensing device with tunable response time [11].

Hereby we aim to describe the response of large-area (up to cm$^2$) CVD graphene to intense single pulses of nanosecond laser as detected by the methods of optical microspectroscopy and scanning electron microscopy (SEM).

## 2. Experimental

Since micromechanical cleavage produces generally too small samples for the experimental arrangement described here, we used two different samples of large-area CVD graphene.

The first sample (later referred to as sample S1) was prepared in home-assembled CVD reactor following the receipt described in [12]. Predominantly monolayer graphene was grown on commercial 25-μm thick polycrystalline copper foils (99.999%, Alfa Aesar). The foils were initially annealed 30 min at 950°C in Ar/H$_2$ (both 99.999%, AS AGA Estonia) flow of 100/120 sccm and then exposed to the flow (40 sccm) of the mixture of 10% CH$_4$ (99.999%, AS AGA Estonia) in Ar at the same temperature for 10 min. Then the sample was cooled at the rate of 15°C/min to room temperature in Ar flow of 100 sccm. Next, the upper sides of copper foils with graphene were covered with PMMA (M~997 000, GPC, Alfa Aesar) in chlorobenzene (Alfa Aesar) solution (20 mg/ml) and the copper was removed with warm FeCl$_3$ (97%, Alfa Aesar) solution (1 mol/l). The PMMA/graphene was washed with deionized water and transferred onto SiO$_2$/Si substrates. PMMA was dissolved by dichloromethane (Alfa Aesar). Finally, the samples were washed in hot acetone (99.5%, Carl Roth GmbH+CO). In addition to monolayer areas (confirmed by Raman spectra) the sample also contained smaller regions of bi- and triple layer graphene which were easily identified visually under optical microscope by contrast differences.

For comparison, similar experiments were conducted on commercial polycrystalline CVD graphene (Graphene Laboratories Inc.), which was also deposited on SiO$_2$/Si substrate. This sample will be referred to as sample S2.

The irradiation of graphene was carried out by focusing the radiation from a tunable pulsed optical parametric oscillator (OPO) Expla NT342/1/UVB (pulse width 5 ns) obliquely on graphene (angle of incidence ~60°) to an elliptical bell-shaped irradiance distribution with FWHM 30 μm x 10 μm (Fig. 1). One UV (266 nm) and one visible (532 nm) wavelength was employed in this study. Only a single pulse was applied with sufficient energy to produce optically visible damage (simultaneously monitored through a microscope). Although the energy of each individual pulse (partly extracted over a quartz plate) was recorded accurately by an energy meter (Ophir Nova II), very accurate determination of damage threshold was complicated due to laser pulse energy fluctuations and some inhomogeneities of the graphene sheet.

Raman spectra were recorded by using a Renishaw inVia micro-Raman spectrometer (spectral resolution 2 $cm^{-1}$) employing the 514 nm line of an argon-ion laser for excitation and a 50x objective for focusing of the laser beam and collecting the backscattered Raman signal.

Microluminescence measurements were carried out on home-assembled setup consisting of Olympus BX41M microscope combined with Andor iXon DU-897D camera for fluorescence imaging and fiber-coupled Andor SR303i spectrometer equipped with Andor Newton DU970P-BV camera for spectral measurements. 532 nm DPSS laser was correspondingly used for either wide-field or point excitation. Dicroic beamsplitter in combination with a band-pass or high-pass filter was used to reject the excitation. The setup also permits employing the OPO at lower intensities for oblique excitation of fluorescence at arbitrary wavelength and confocal detection [13].

X-ray photoelectron spectroscopy (XPS) measurements were performed in UHV chamber with a SCIENTA SES-100 spectrometer using the Mg Kα X-ray source (photon energy 1253.6 eV, FWHM 0.7 eV). The error of measured absolute energies is less than 0.1 eV.

High-resolution scanning electron microscopic (HR-SEM) images were acquired by using Helios[TM] NanoLab 600 (FEI) system operated at probe electron energies of 1–5 keV and current of 21 pA. Secondary electron images of the sample were acquired with TLD (Thru-the-Lens Detector).

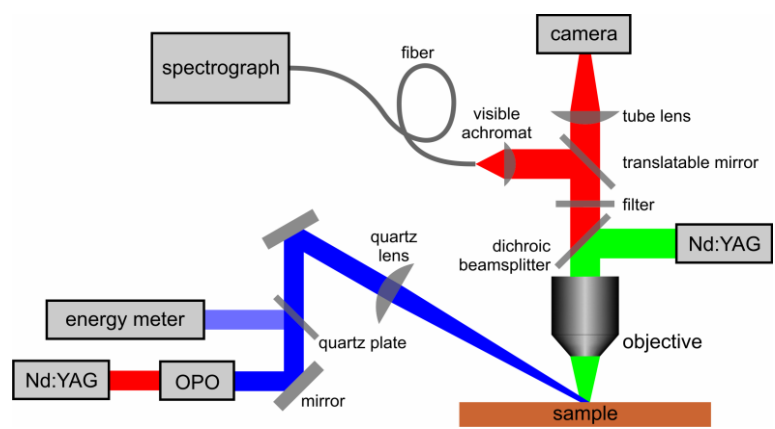

Figure 1. Experimental setup used for irradiation of graphene and for fluorescence microspectroscopic measurements. Additional optics inserted for wide-field excitation is omitted for clarity.

## 3. Results and discussion

Laser-induced formation of optically clearly visible damage area in the graphene layer was observed typically for peak energy densities above ~200 mJ/cm$^2$. Exceptionally, the home-grown sample S1 appeared much more sensitive to UV pulses with damage threshold as low as 25 mJ/cm$^2$. A possible explanation is that the damage mechanism involves absorption of UV radiation by some adsorbed species which are present at higher concentration on the CVD samples grown in our lab. Moreover, it is known that the absorption of graphene itself becomes markedly stronger around 270 nm [2]. This fact has to be combined with the possibility that due to the peculiarities of the transfer process applied to commercial graphene, it has much better thermal contact with substrate leading to more efficient dissipation of absorbed energy. This is partially suggested by the fact that under UV pulses the visible damage in the commercial graphene sheet (S2) appears nearly at the same intensities as the damage in the underlying SiO$_2$/Si substrate. Therefore the details of the damage process under UV and visible irradiation may be slightly different and influenced by the presence of substrate.

In order to check the possibility of chemical difference between the samples, XPS measurements were carried out. Due to the large beam size of the used XPS apparatus only untreated graphene areas could be characterized by this method. XPS scans of the C 1s core level peak are depicted in Fig. 2. It was possible to deconvolute the spectra into four Gaussian components. The strongest peak centered at 284.8 eV corresponds to C 1s in sp2 configuration (i.e. graphene). The remaining peaks at 285.6, 286.4 and 288.5 eV belong most probably to C–OH, C–O and C=O bonds [14]. Differences in the oxidation level are relatively small (some interplay between the amount of single and double C-O bonds). It appears that the peak corresponding to C–OH is almost completely missing in the commercial sample (S2). Therefore is seems that the tentative candidate influencing the damage threshold under UV irradiation might be the OH-groups which is also in accordance with the fact that no special heat treatment has been applied to the as-grown samples produced in our lab. Further comparative measurements with differently annealed samples will be necessary to confirm this assessment.

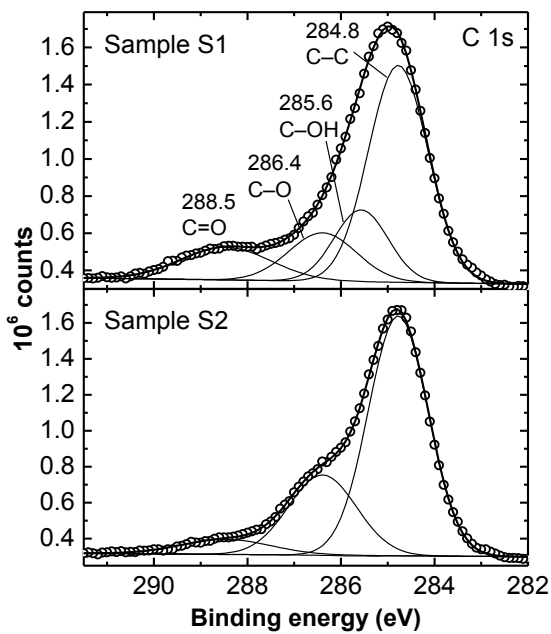

Figure 2. High-resolution XPS scans of the C 1s core level peak with a decomposition into Gaussian components.

It is remarkable that the estimated damage threshold is quite close to the single shot damage threshold for femtosecond lasers. Currie at al. [9] and Yoo et al. [10] have estimated 57 mJ/cm$^2$ and 98 mJ/cm$^2$, respectively, as the ablation thresholds for 50 fs and 100 fs pulses at 800 nm. Roberts et al. [8] have reported slightly higher value of 200 mJ/cm$^2$ for pulse lengths of 50 fs…1.6 ps. Our result suggests that the relaxation time of the absorbed energy determining the ablation threshold must be in the range of nanoseconds or longer so that even for nanosecond pulses the total pulse energy rather than peak power is critical for the onset of damage.

Typical SEM image of the damaged area is shown in Fig. 3a. In the central area the regular graphene network is completely destroyed (ablated). Bordering of this ablated area is a proportionally sized stripe consisting of fragments of graphene. The width of this stripe is considerably smaller than the central area which suggests that the energy density threshold for ablating of graphene is still relatively well-defined.

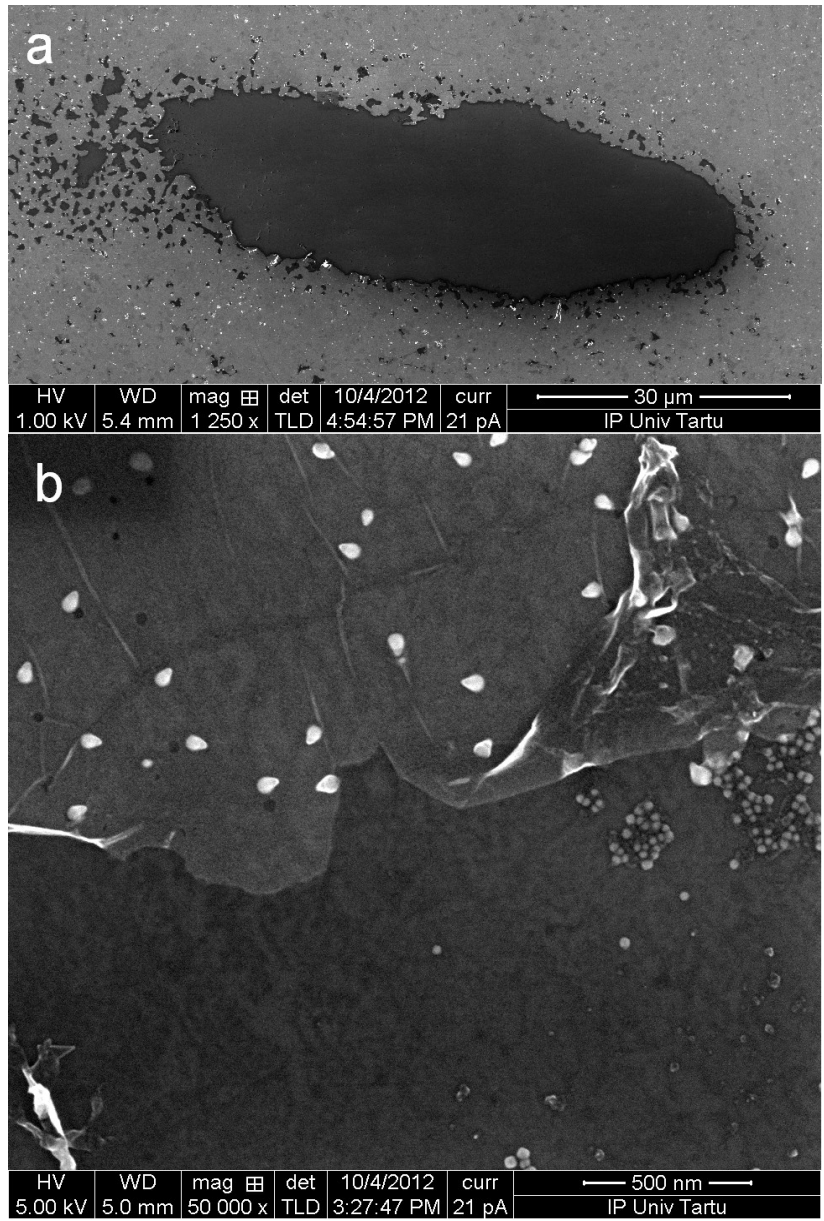

Figure 3. (a) SEM micrograph of single laser pulse (360 mJ/cm$^2$ at 532 nm) induced damage on sample S2. The strongly fragmented area on the left is due to a slight asymmetry of the laser pulse irradiance distribution. (b) Partial folding of graphene observed at the border of ablated area on sample S1 created with 160 mJ/cm$^2$ laser pulse at 266 nm.

The nature of the damage was further assessed with micro-Raman mapping. First we note that the Raman spectrum of both pristine samples exposed nearly Lorentzian shaped 2D peak and high $I_{2D}/I_G$ ratio indicating single graphene layer [15]. Figure 4 depicts the representative behavior of Raman spectrum along the line crossing the perimeter of ablated area. Three trends are immediately evident when moving from the undamaged region into the damaged one.

I. Intensity of the defect-activated D-band becomes systematically stronger compared to the G band. The ratio $I_D/I_G$ of the intensities of the D and G bands achieves in some cases a value as high as 2.1 at the boundary of the damaged area. Some contribution to the appearance of the D peak may come from the presence of visible edges as evident from SEM image (corresponding $I_D/I_G$ may be up to ~0.5 [17]). But in general we can expect more finely distributed damage of graphene due to laser induced bond breaking and formation of nanocrystalline $sp^2$ network. Several studies exist which have quantified the amount of graphene disorder based on $I_D/I_G$. The maximal $I_D/I_G$ value observed by us can be correlated to the average $sp^2$ cluster size of ~7 nm according to the formula given in Ref. [18] or the average distance between point defects ~8 nm according to Ref. [19].

II. In some cases the absolute intensity of the G-band is temporarily enhanced in the proximity of the perimeter indicating the formation of a narrow stripe of graphene bilayer. Such bilayer is also visible in the optical image (Fig. 4a). SEM image (Fig. 2b) indicates random folding of graphene along the boundary of the damage. More regular folding pattern was previously reported after single shot irradiation with tightly focused femtosecond laser where it was also demonstrated that the presence of substrate is essential for appearance of the folds [10]. Here we observed graphene folds mainly after UV irradiation of sample S1 suggesting that the phenomenon may have similar origin as the decreased damage threshold of the sample.

III. In the central damaged region the signal becomes faint but remains detectable. The general shape of the Raman spectrum of the residues matches rather well chemically derived graphene oxide [20,21] as well as an intermediate stage of defective graphene achieved with oxygen plasma treatment [22].

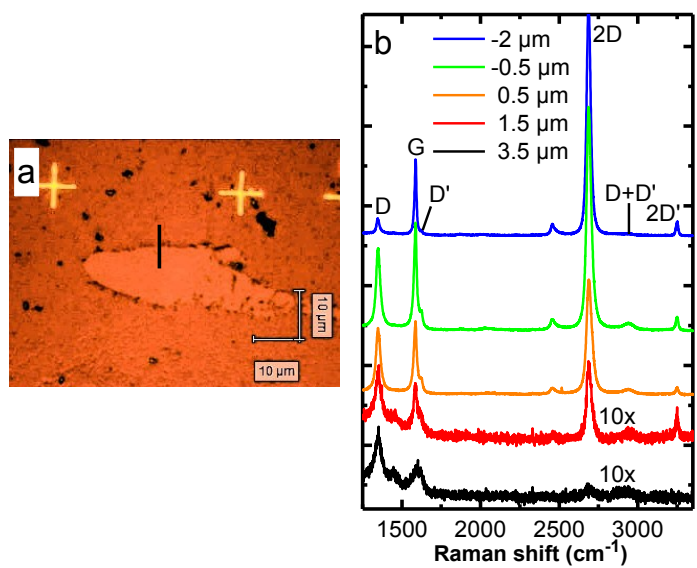

Figure 4. (a) Optical micrograph of damage created in sample S1 by 160 mJ/cm$^2$ laser pulse at 266 nm. (b) Raman spectra mapped along the path shown by black vertical line in (a). The very weak feature at 1450 cm$^{-1}$ may be due to the traces of PMMA used in the transfer process of graphene sheet from Cu-foil to SiO$_2$/Si substrate [16].

Next the fluorescence microspectroscopy of irradiated graphene was carried out. In most cases a clear (though spatially not very homogeneous) fluorescence signal was detected from the irradiated areas (Fig. 5b). The low contrast ratio compared to non-irradiated areas indicates that the fluorescence is apparently much weaker than has been previously obtained from e.g. oxygen plasma treated graphene [22]. Yet, the signal is surprisingly strong considering that only traces of carbon (as detected by Raman spectrum) are present. The spectral shape of the signal matches quite well the fluorescence obtained from oxygen plasma treated graphene [22] and graphene oxide [21]. The assignment of this fluorescence to the residues of graphene is therefore in accordance with the Raman spectrum.

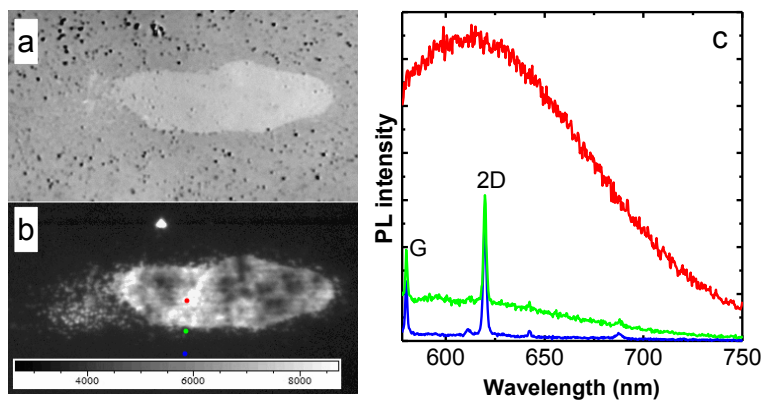

Figure 5. (a) Bright-field optical micrograph of irradiated area of graphene on sample S2 (532 nm, 360 mJ/cm$^2$). (b) Fluorescence image of the same area under 532 nm excitation viewed through 630–710 nm band-pass filter. (c) Fluorescence spectra acquired from the corresponding points marked in (b).

## 4. Conclusions

We have found that single laser pulses with 5 ns duration produce visible damage in monolayer CVD graphene above threshold energy density of ~200 mJ/cm$^2$. Under UV irradiation, however, only commercial sample was capable to resist such intensities whereas home-gown sample developed damage already under an order of magnitude lower irradiance. UV irradiation of the latter sample also leads to the formation of folded graphene at the boundary of the central ablated area. These peculiarities may appear due to the presence of some additional impurities in the home-grown sample (e.g. OH-groups) and/or the lack of thermal contact between the transferred graphene sheet and the substrate suppressing effective dissipation of absorbed energy. The residues in the virtually graphene-free ablated areas expose surprisingly strong fluorescence spectrally matching that reported for defective or chemically modified graphene structures (oxidized graphene) obtained by other methods. Further research is necessary with more refined treatment at lower fluences to determine whether defect engineering without visible damaging of graphene is possible with nanosecond lasers.

**Acknowledgements**

The authors are grateful to Dr. Raivo Jaaniso for fruitful discussion. This research was carried out with the financial support of Estonian Science Foundation (Grants 9283, SF0180058s07 and SF0180046s07) and by the European Union through the European Social Fund (Grant MTT1) and European Regional Development Fund (Centers of Excellences "Mesosystems: Theory and Applications", TK114 and "High-technology Materials for Sustainable Development", TK117T).

**References**

[1] A.K. Geim and K.S. Novoselov, The rise of graphene, Nature Mat. 6 (2007), 183–191.

[2] F. Bonaccorso, Z. Sun, T. Hasan, and A.C. Ferrari, Graphene photonics and optoelectronics, Nature Phot. 4 (2011), 611–622.

[3] Y. Tao, B. Varghese, M. Jaiswal, S. Wang, Z. Zhang, B. Oezyilmaz, K. Loh, E. Tok, and C. Sow, Localized insulator-conductor transformation of graphene oxide thin films via focused laser beam irradiation, Appl. Phys. A (2012), 1–9.

[4] L. Guo et al., Bandgap tailoring and synchronous microdevices patterning of graphene oxides, J. Phys. Chem. C 116 (2012), 3594–3599.

[5] Y.-L. Zhang, L. Guo, S. Wei, Y. He, H. Xia, Q. Chen, H.-B. Sun, and F.-S. Xiao, Direct imprinting of microcircuits on graphene oxides film by femtosecond laser reduction, Nano Today 5 (2010), 15–20.

[6] J.-N. Wang, R.-Q. Shao, Y.-L. Zhang, L. Guo, H.-B. Jiang, D.-X. Lu and H.-B. Sun, Biomimetic graphene surfaces with superhydrophobicity and iridescence, Chem. Asian J. 7 (2012), 301–304.

[7] B. Krauss, T. Lohmann, D.H. Chae, M. Haluska, K. von Klitzing, and J.H. Smet, Laser-induced disassembly of a graphene single crystal into a nanocrystalline network, Phys. Rev. B 79 (2009), 165428.

[8] A. Roberts, D. Cormode, C. Reynolds, T. Newhouse-Illige, B.J. LeRoy, and A.S. Sandhu, Response of graphene to femtosecond high-intensity laser irradiation, Appl. Phys. Lett. 99 (2011), 051912.

[9] M. Currie, J.D. Caldwell, F.J. Bezares, J. Robinson, T. Anderson, H. Chun, and M. Tadjer, Quantifying pulsed laser induced damage to graphene, Appl. Phys. Lett. 99 (2011), 211909.

[10] J.-H. Yoo, J.B. In, J.B. Park, H. Jeon, and C.P. Grigoropoulos, Graphene folds by femtosecond laser ablation, Appl. Phys. Lett. 100 (2012), 233124.

[11] L. Guo, et al., Two-beam-laser interference mediated reduction, patterning and nanostructuring of graphene oxide for the production of a flexible humidity sensing device, Carbon 50 (2012), 1667–1673.

[12] A.M. van der Zande, R.A. Barton, J.S. Alden, C.S. Ruiz-Vargas, W. S. Whitney, P. H. Q. Pham, J. Park, J. M. Parpia, H.G. Craighead and P.L. McEuen, Large-scale arrays of single-layer graphene resonators, Nano Lett. 10 (2010), 4869–4873.

[13] S. I. Omelkov, V. Kiisk, I. Sildos, M. Kirm, V. Nagirnyi, V. A. Pustovarov, L. I. Isaenko, S. I. Lobanov, The luminescence microspectroscopy of $Pr^{3+}$-doped $LiBaAlF_6$ and $Ba_3Al_2F_{12}$ crystals, Radiat. Meas. (accepted).

[14] R. Hawaldar et al., Large-area high-throughput synthesis of monolayer graphene sheet by hot filament thermal chemical vapor deposition, Scientific Rep. 2 (2012), article no. 262.

[15] A.C. Ferrari, et al., Raman spectrum of graphene and graphene layers, Phys. Rev. Lett. 97 (2006), 187401.

[16] L. Jiao, L. Zhang, X. Wang, G. Diankov, and H. Dai, Narrow graphene nanoribbons from carbon nanotubes, Nature 458 (2009), 877–880.

[17] C. Casiraghi, A. Hartschuh, H. Qian, S. Piscanec, C. Georgi, A. Fasoli, K.S. Novoselov, D.M. Basko, and A.C. Ferrari, Raman spectroscopy of graphene edges, Nano Letters 9 (2009), 1433–1441.

[18] M.A. Pimenta, G. Dresselhaus, M.S. Dresselhaus, L.G. Cançado, A. Jorio, and R. Saito, Studying disorder in graphite-based systems by Raman spectroscopy, Phys. Chem. Chem. Phys. 9 (2007), 1276–1290.

[19] L.G. Cançado, et al., Quantifying defects in graphene via Raman spectroscopy at different excitation energies, Nano Letters 11 (2011), 3190-3196.

[20] A. Wei, J. Wang, Q. Long, X. Liu, X. Li, X. Dong, and W. Huang, Synthesis of high-performance graphene nanosheets by thermal reduction of graphene oxide, Mat. Res. Bull. 46 (2011), 2131–2134.

[21] T.V. Cuong, V.H. Pham, Q.T. Tran, S.H. Hahn, J.S. Chung, E.W. Shin, and E.J. Kim, Photoluminescence and Raman studies of graphene thin films prepared by reduction of graphene oxide, Mat. Lett. 64 (2010), 399–401.

[22] T. Gokus, R.R. Nair, A. Bonetti, M. Böhmler, A. Lombardo, K.S. Novoselov, A.K. Geim, A.C. Ferrari, and A. Hartschuh, Making Graphene Luminescent by Oxygen Plasma Treatment, ACS Nano 3 (2009), 3963–3968.